\documentclass[%
 reprint,
 amsmath,amssymb,
 aps,
]{revtex4-1}

\usepackage{graphicx}% Include figure files
\usepackage{dcolumn}% Align table columns on decimal point
\usepackage{bm}% bold math
\begin{document}

%\preprint{APS/123-QED}

\title{The Local Universe from Calar Alto (LUCA).\newline 
Unraveling from 3D spectroscopy the fine structure of galaxies in the Local Volume}% Force line breaks with \\
%\thanks{Expunged/liquidated/deceased September $\mathrm{8^{th}}$, 2019}%

\author{Francisco Prada}
 \altaffiliation[]{f.prada@csic.es}%Lines break automatically or can be forced with \\
\author{Enrique P\'erez}%
\affiliation{%
Instituto de Astrof\'\i sica de Andaluc\'\i a, CSIC\\
Glorieta de las Astronom\'\i a s/n, E-18008 Granada, Spain
}%

%\collaboration{LUCA Collaboration}%\noaffiliation

\author{and the LUCA collaboration}

\date{\today}% It is always \today, today,

\begin{abstract}

LUCA (Local Universe from Calar Alto) was conceived as a new generation science program for the Calar Alto Observatory. It proposed the construction of a large Integral Field Unit (IFU) spectrograph with six thousands optical fibers (IFU-6000) at the CAHA 3.5-m telescope to map our universe neighborhood in 3D with an unprecedented spatial resolution. Two galaxy samples were defined to map the local universe: (i) 102 galaxies in the Local Volume, out to 11 Mpc, and (ii) 218 galaxies in the Virgo cluster. A complementary project was developed to map the three largest galaxies in the local universe, M31, M33, and M101 with the Schmidt telescope. In this white paper we describe the LUCA Project science justification and survey strategy, as well as a technical description of the IFU-6000 instrument, its performance and design.

\end{abstract}

\pacs{Valid PACS appear here}% PACS, the Physics and Astronomy
\maketitle

%\tableofcontents

%---------------------------------------------------------------------------------------------------------------------------------------------------------------------------
%---------------------------------------------------------------------------------------------------------------------------------------------------------------------------
\section{\label{sec:intro}Introduction}

LUCA (Local Universe from Calar Alto) was proposed as a new generation science program for the Calar Alto Observatory (CAHA) 3.5-m telescope. The LUCA Project faces the construction of a large Integral Field Unit (IFU) spectrograph with six thousands optical fibers (named IFU-6000), that will allow astronomers to map our universe neighborhood in 3D with an unprecedented spatial resolution. LUCA will be able to observe large pieces of the sky generating a massive quantity of spectral data cubes in the optical range, aiming to unravel physical processes at small enough scale to study how star formation affects the formation and evolution of galaxies in our Local Universe.

The project, conceived and led by the Instituto de Astrof\'\i sica de Andaluc\'\i a (IAA-CSIC), was selected by the CAHA's Advisory Committee to finance its Feasibility Study in response to the CAHA announcement released on May 11, 2018 for new instrumentation. The feasibility study phase extended over 9 months during the period October 1, 2018, to July 12, 2019. A total of 243,300 euro was required for its accomplishment. 

This document summarises the Feasibility Study of the LUCA Project, and it references in the relevant sections the detailed work done for each of the instrument systems. The work included in this report has been performed by a team of engineers and scientists at the IAA-CSIC in collaboration with the institutions and consultants listed below (see Sec.~\ref{sec:acknow}). We provide science specifications and requirements, science overview and survey strategy, a technical description of the instrument proposal, performances and designs. Conclusions and remarks are given in Sec.~\ref{sec:conclus}.

%---------------------------------------------------------------------------------------------------------------------------------------------------------------------------
%---------------------------------------------------------------------------------------------------------------------------------------------------------------------------
\section{\label{sec:lucascience}LUCA Science}

Hierarchical clustering cosmology has been successful to explain the large-scale structure of the universe, which is formed by a complex web of clusters, filaments, and voids. However, at small and intermediate scales, there is a significant number of problems to explain how galaxies form and how they evolve depending on their environment. Recently, high spatial resolution cosmological simulations of galaxy formation and evolution have started to be developed. These models require the implementation of accurate phenomenological prescriptions for the sub-grid physics governing galaxy evolution, in particular those that link the star formation processes with the galactic feedback, which occur at few tens parsec scales.  

Galaxies are a complex mix of baryonic (stars, gas, and dust) and dark matter, and radiation, spatially distributed in galaxy components (bulge, thin and thick disks, halo, etc) at kpc scales. However, the physical processes that lead to star formation, and the modes by which this activity couples to the broader environment where galaxies flow, occur on smaller scales, of a few tens of parsecs. Here, we propose to study the Local Universe from these small parsec scales to the scales of the nearest galaxy cluster, i. e., Virgo. This will improve by more than an order of magnitude the spatial resolution of current Integral Field Spectroscopy (IFS) galaxy surveys (e.g. CALIFA, MaNGA). 

As defined by the IAU, based on current resolution of individual stars, the Local Universe comprises a sphere of radius 15 Mpc centered in the Local Group, and it includes the Local Volume and the Virgo cluster. 

Galaxies in the Local Universe offer us the unique opportunity to map the kinematics, physical and chemical properties of the stellar populations, the interstellar medium, and dark matter at spatial scales from a few pc, as in M31 or M33, to less than 100 pc in the Virgo cluster. This is the range of spatial scales needed to constrain the sub-grid physics in models of galaxy formation and evolution. 

Furthermore, because over this Local Volume all the galaxies are considered to be at roughly the same cosmic epoch, but the local density field varies by a factor of $\sim100$, the large-scale evolutionary effects are due more to the underlying density than to any variation in time. Thus, LUCA will enable us to map the bulge, disk, and halo properties with unprecedented high spatial resolution, and to study the effects of the environment on these galaxy components at different levels of field density, isolated galaxies, galaxy groups, and the nearest galaxy cluster.

In addition to this main global science case, many other science cases can be carried out with the proposed instrumentation, including extended stellar and nebular targets in the Milky Way.

\begin{center}
\begin{table}
\centering
\begin{tabular}{c | c } 
\hline
 Science requirement & Parameter value  \\ 
 \hline
Telescope & CAHA 3.5m  \\ 
IFU FoV    & 9 arcmin$^2$ \\
Spatial sampling & $2.5''$ \\
\# spaxels & 6000 \\
Fiber core on CCD & $< 4$ pixels \\
On sky FoV effective coverage & $> 90$\% \\
Fiber cable length & $< 40$m \\
Spectral resolution @ 500nm & 2000 \\
Wavelength range & 360--700 nm \\
Fiber cross-talk & $\leq5\%$ \\
\hline
\end{tabular}
\caption{Summary of main science technical requirements}
\label{table:1}
\end{table}
\end{center}

\subsection{\label{sec:level2}LUCA Sample}

Following the IAU definition of the Local Universe, LUCA is a survey of nearby galaxies selected to be in the Local Volume and in the Virgo cluster. Thus, the survey contains two galaxy samples.

\hfill\break

{\bf First sample: Local Volume}

A selection of galaxies from the complete sample of galaxies in the Local Volume, up to 11 Mpc, taken from Klypin et al. (2015), that are visible from CAHA and are more luminous than $\rm M_B < -16$. This selection includes all spirals and early-type galaxies of the Local Volume with stellar mass above $10^9$ M$\odot$. Excluding M31, M33, and M101, that are the largest members of this sample (they are the specific targets of an independent proposal for the CAHA Schmidt telescope IFU project, with documentation already submitted to the CAHA Director, and also rejected), the galaxies have a size ($\rm a_{26}$, the isophotal diameter at a B mag surface brightness $\rm SB_B=26$ mag/arcsec$^2$) that ranges from 1 to 30 arcmin, with mean 7.4 arcmin and median 5.6 arcmin. 

Thus, the LV sample is formed by 102 galaxies at a mean distance of 6.7 Mpc, providing a spatial scale of 32 pc/arcsec, and having a mean diameter size of 7.4 arcmin. 
 
%% FIGURE:
\begin{figure}
\centering
\includegraphics[width=0.5\textwidth]{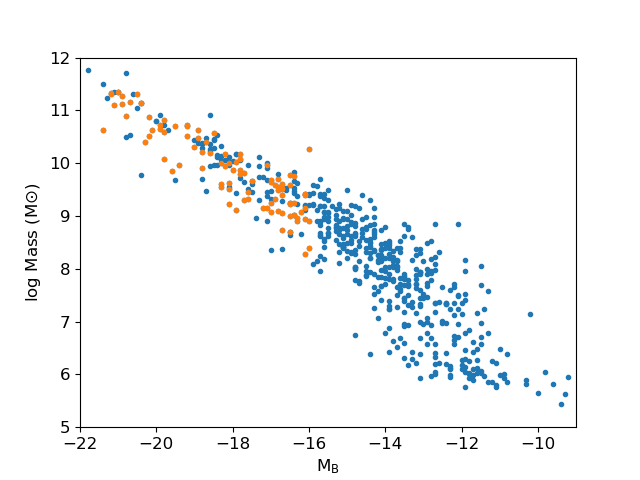}
\caption{
Blue points represent all the 603 LV galaxies (Klypin et al. 2015) with a measured $\rm M_B$ and estimated dynamical mass. In orange the 102 galaxies of the LUCA LV sample, with $\rm M_B < -16$ and visible from Calar Alto.
}
\label{fig:LV_Mass_Mb}
\end{figure}

%% FIGURE:
\begin{figure}
\centering
\includegraphics[width=0.5\textwidth]{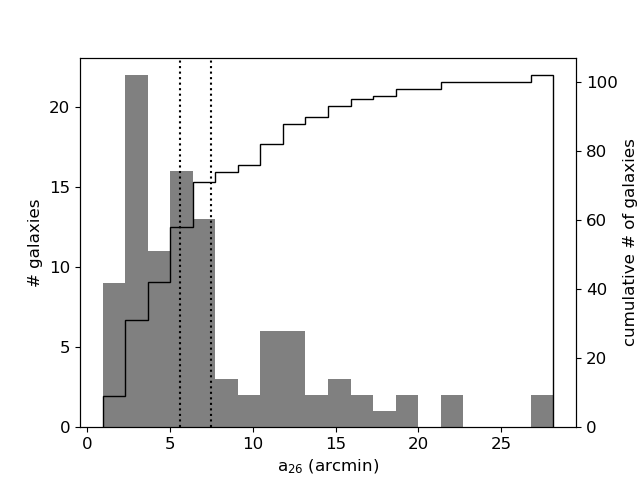}
\caption{
Distribution of galaxy sizes ($\rm a_{26}$, the isophotal diameter at a surface brightness $\rm SB_B=26$ mag/arcsec$^2$) for the Local Volume sample. The gray scale filled histogram corresponds to the left y-axis, while the black hollow line shows the cumulative distribution (right hand y-axis), to a total of 102 galaxies. Vertical dotted lines indicate mean ($7.4'$) and median ($5.6'$) values.
}
\label{fig:LV_a26_histo}
\end{figure}

%% FIGURE:
\begin{figure}
\centering
\includegraphics[width=0.5\textwidth]{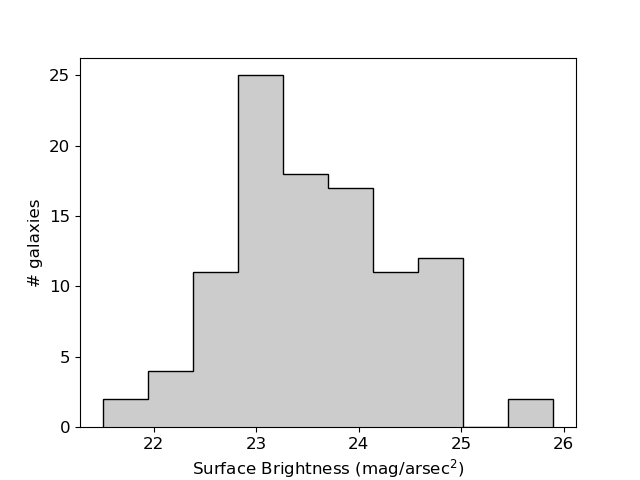}
\caption{
Distribution of surface brightness for the 102 galaxies in the LUCA LV sample.
}
\label{fig:LV_SB}
\end{figure}

%% FIGURE:
\begin{figure}
\centering
\includegraphics[width=0.5\textwidth]{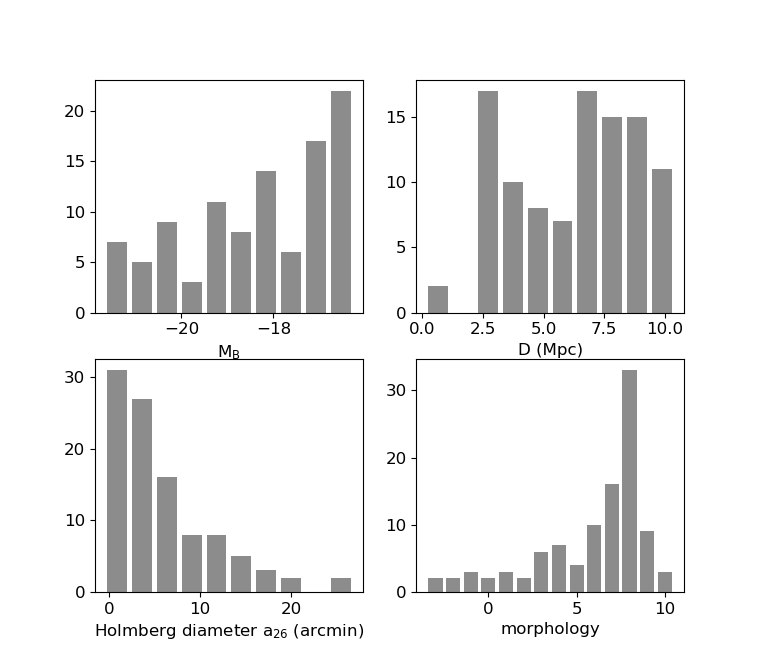}
\caption{
Distribution of magnitude, distance, size, and morphology for the 102 galaxies in the LUCA LV sample.
}
\label{fig:LV_histos}
\end{figure}

\hfill\break

{\bf Second sample: Virgo cluster}

We have done a selection of 218 galaxies from the Virgo Cluster Catalog (VCC, Binggeli, Sandage \& Tammann 1985) more luminous than $\rm M_g < -17$ and visible from Calar Alto. This sub-sample includes galaxies of all morphological types and are spatially well distributed within the cluster, covering its main filamentary structures. The sample is complete for galaxies with stellar mass $>10^9$ M$\odot$; it provides a spatial scale of 80 pc/arcsec and a mean size of 2.4 arcmin.

We will seek ways to provide synergy with current and upcoming surveys of the Local Universe at different wavelengths, such as those provided by Spitzer FIR (LVL survey), SKA and ngVLA (SWG2) HI, and other optical surveys (HST/LEGUS, J-PLUS/J-PAS).

To our knowledge, no other project is planned to map with IFS such an extent of the Local Volume. The closest competitor beyond 2020 will be SDSS-V, which includes as one of its three main surveys the Local Volume Mapper concentrating on the MW and LMC. For M31, M33 and other galaxies out to 5 Mpc they will have a sparse IFS sampling: ``Statistical samples of H II regions (green) observed at 20 pc resolution across M31, and at $\sim50$ pc resolution across other nearby galaxies" (cf. arXiv1711.03234). In the North they will have only $\sim2000$ fibers, compared to the $\sim6000$ for LUCA. We had planned to reach the sky at about the same time.

%% FIGURE:
\begin{figure}
\centering
\includegraphics[width=0.5\textwidth]{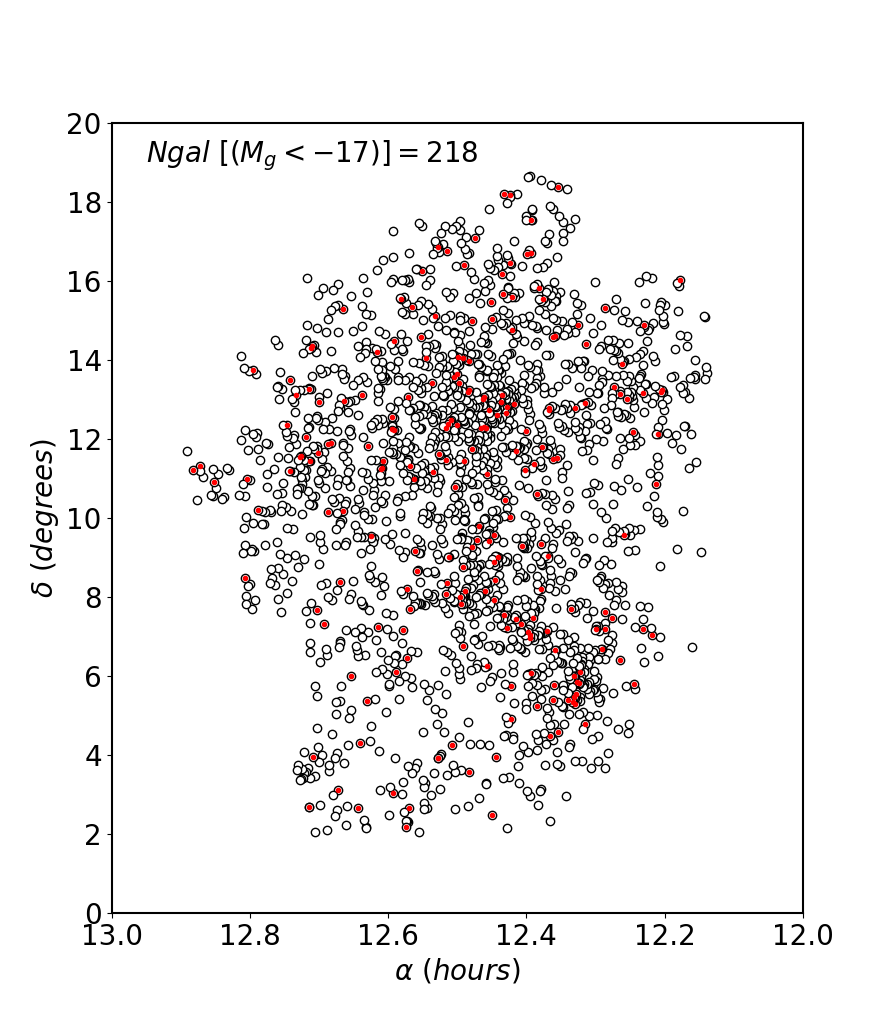}
\caption{
Spatial distribution of galaxies in the Virgo Cluster Calatog. In red the galaxies selected in the LUCA VCC sample, with $\rm M_g < -17$ and visible from Calar Alto. 
}
\label{fig:VCC_ra_dec}
\end{figure}

%% FIGURE:
\begin{figure}
\centering
\includegraphics[width=0.5\textwidth]{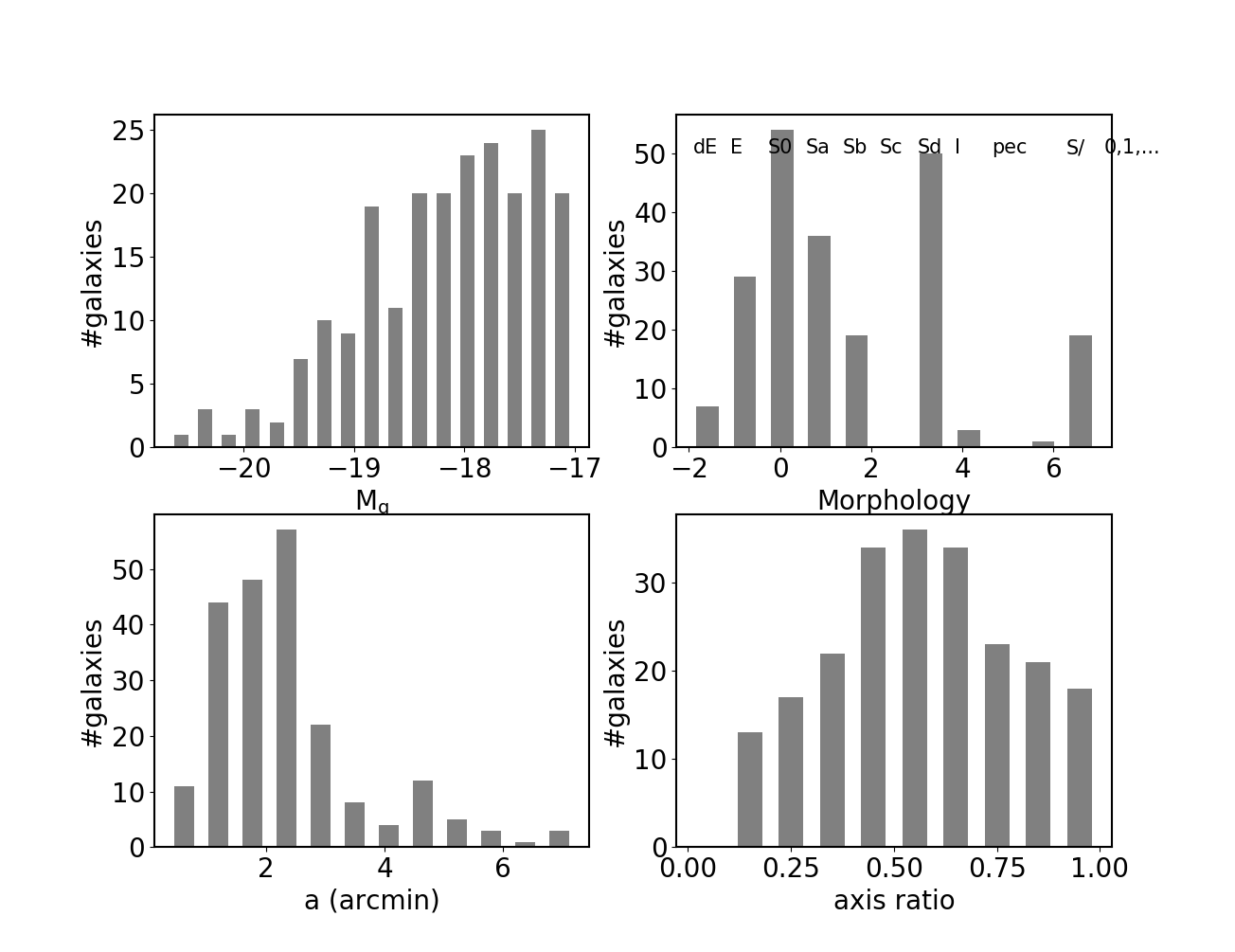}
\caption{
Distribution of galaxy properties in the LUCA VCC sample. Mg is the absolute magnitude in the SDSS g-band, the morphological type, the semi-major axis size (in arcmin) and the galaxies axis ratio (0.1 for edge-on, 1 for face-on).  
}
\label{fig:VCC_histos}
\end{figure}

%---------------------------------------------------------------------------------------------------------------------------------------------------------------------------
%---------------------------------------------------------------------------------------------------------------------------------------------------------------------------
\section{\label{sec:surveystrateg}LUCA Survey Strategy}

The LV sample used in the study of Klypin et al. (2015) is drawn from the work of Karachentsev et al. (2013, AJ, 145, 101; http://www.sao.ru/lv/lvgdb/) with the following selection criteria,

\begin{itemize}
\item Range of declination so that the target is visible at least one hour, equivalent to a target maximum altitude of at least 50 degrees. This results in $\rm -2.77 < dec < 77.23$
\item The Local Volume is defined as the universe within a radius of $\rm D \leq 11$ Mpc from the Milky Way. Data compiled by Klypin et al. (2015, MNRAS, 454, 1789).
\item Limiting absolute blue magnitudes $\rm M_B \leq -16$ 
\item Measured galaxy major-axis smaller that $30'$ to exclude the very large galaxies (M31, M33, and M101) that could be done with an IFU at the CAHA Schmidt telescope. Thus $\rm  a_{26} < 30'$.
\end{itemize}

\hfill\break

\begin{figure}
\centering
\includegraphics[width=0.5\textwidth]{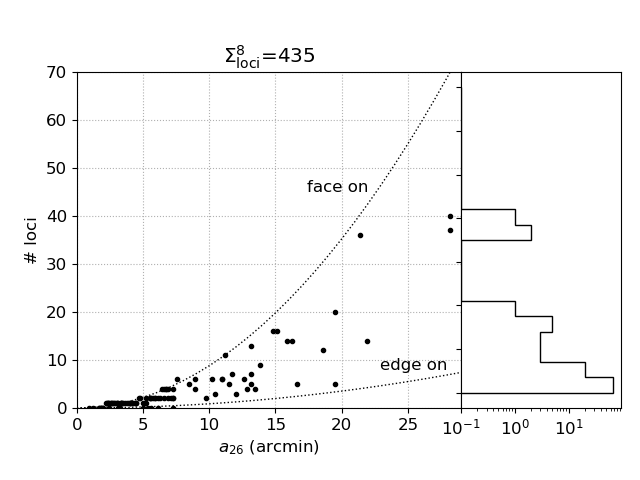}
\caption{
Number of loci per galaxy in the LUCA LV sample, for a total of 435 loci for the 102 galaxies. 
}
\label{fig:LV_loci_a26_modelo_8}
\end{figure}

Figure~\ref{fig:LV_loci_a26_modelo_8} shows the number of loci (we use the term locus / loci as meaning pointing / pointings) required to cover each galaxy, with the full 8-spectrographs system of $\rm FoV = 9.1\ arcmin^2$. The spread along the vertical axis for a given $\rm a_{26}$ is produced by the inclination of the galaxy; for the same major-axis, edge-on galaxies require less loci than face-on galaxies. This is illustrated with the dotted lines, that represent the limiting cases.

%% FIGURE:
\begin{figure}
\centering
\includegraphics[width=0.5\textwidth]{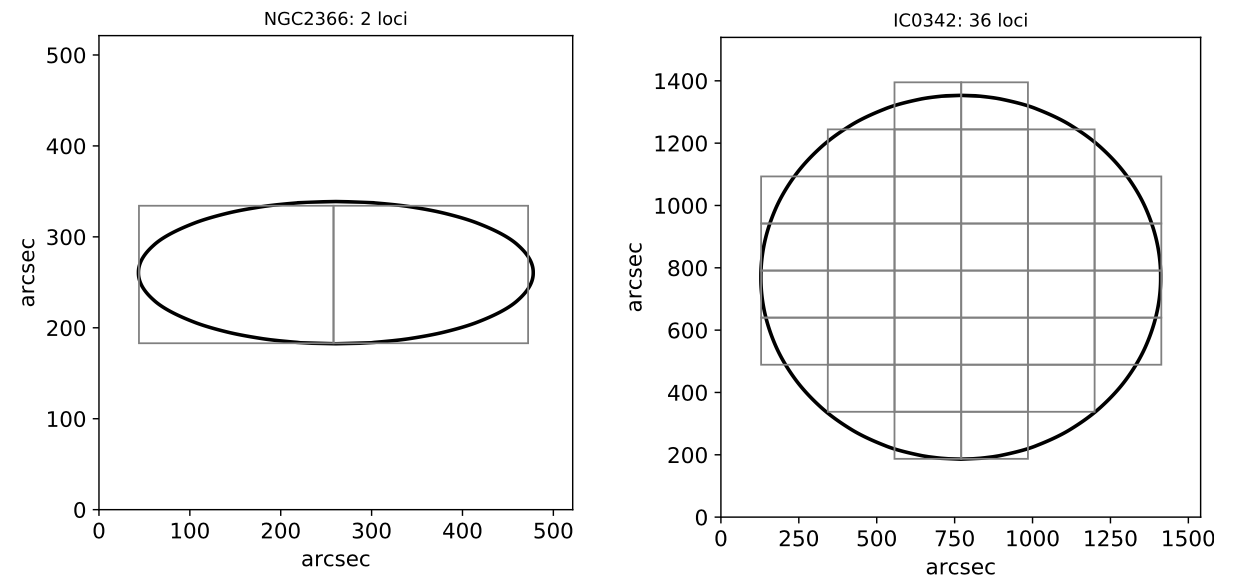}
\caption{
Examples of low and high number of pointings needed to completely map a galaxy with the 8-spectrographs system. Each box represents the IFU $\rm FoV = 9.1\ arcmin^2$. The ellipse represents the galaxy by its semi-major axis and axis ratio.
}
\label{fig:fill_examples}
\end{figure}

%% FIGURE:

The VCC sample is drawn from many works, originally by Binggeli et al (1985, AJ,90, 1681), and more recently by the Next Generation Virgo Cluster Survey (NGVS) (Ferrarese et al 2012 ApJS, 200, 4, and series of papers). We have obtained the data tables from Simbad. Our selection criteria is as follows,

\begin{itemize}
\item Range of declination so that the target is visible at least one hour, equivalent to a target maximum altitude of at least $50º$. This results in $\rm -2.77 < dec < 77.23$,
\item Limiting absolute blue magnitude $\rm M_g \leq -17$. 
\end{itemize}

%% FIGURE:
\begin{figure}
\centering
\includegraphics[width=0.5\textwidth]{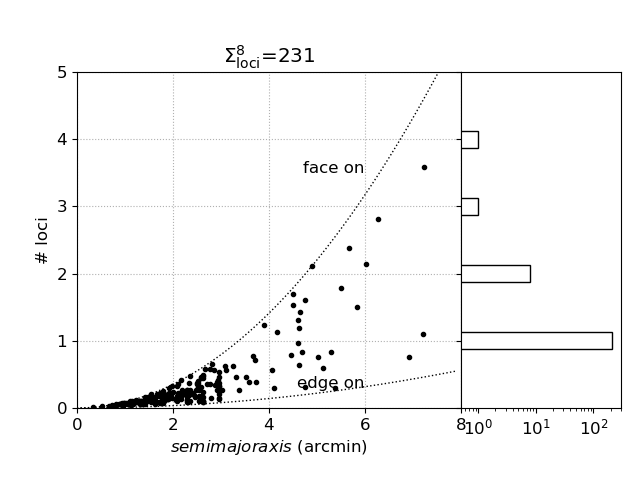}
\caption{
Number of pointings per galaxy in the LUCA VCC sample, for a total of 231 loci required for the 218 galaxies. Notice that the points with fractional values of loci are in fact upgraded to their ceiling value, as shown in the histogram on the right.
}
\label{fig:VCC_loci_a26_modelo_8}
\end{figure}

We have found a linear relation between the total area of loci needed to cover a galaxy and the area of the galaxy, such that

\begin{equation}
\rm \#loci \times area_{IFU} = 0.255 \times area_{gal} .
\end{equation}

From this relation, the number of loci is given by

\begin{equation}
\rm \#loci = 0.255 \times area_{gal} / area_{IFU} .
\end{equation}

This relationship is independent of the number of spectrographs used, as illustrated in the figure for the whole IFU of 8 spectrographs and for the same system when only the first spectrograph has been commissioned and is used.

Notice that although this relationship can give a fractional number of loci, meaning that the galaxy is smaller than the IFU F.o.V, in practice values less than 1 are upgraded to 1, and in general fractional loci are upgraded to their ceiling value.

%% FIGURE:
\begin{figure}
\centering
\includegraphics[width=0.5\textwidth]{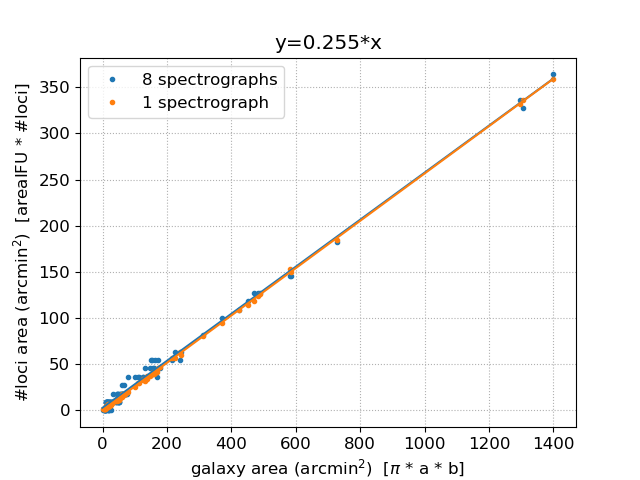}
\caption{
Linear relationship between the area of the galaxy and the area of loci needed to cover the galaxy. 
}
\label{fig:gxArea_nifusArea}
\end{figure}

\begin{center}
\begin{table}
\centering
\begin{tabular}{c | c | c | c | c | c |} 
\hline
 sample & constraint & $\#$ galaxies & $\#$ loci & 8 sp 9.1 & 1 sp 1.1  \\ 
 \hline
Local Volume & $\rm M_B\leq-16$ & 102 & 435 & 53 & 3 \\ 
Virgo Cluster  & $\rm M_g\leq-17$ & 218 & 231 & 208 & 114 \\
\hline
\end{tabular}
\caption{Summary table with the total number of galaxies and pointings for the two samples, and the number of galaxies that can be observed with just one pointing, with 8 spectrographs ($\rm FoV = 9.1\ arcmin^2$) and with 1 spectrograph ($\rm FoV = 1.1\ arcmin^2$).}
\label{table:2}
\end{table}
\end{center}

Table~\ref{table:2} shows, for the two samples, the number of galaxies, the total number of loci necessary to observe them with the complete 8-spectrograph system (FoV 9.1 arcmin$^2$) and the number of galaxies that can be observed with just one pointing (one loci, $\rm FoV = 1.1\ arcmin^2$) either with the full 8-spectrograph system or with only the first spectrograph. Thus, we can see that if the first spectrograph is commissioned, about half of the Virgo sample can be efficiently observed with just one pointing per galaxy, while only three galaxies of the LV sample could be observed with just one pointing and one spectrograph.

The total number of observing hours is computed as $\rm hours = \Sigma(loci) \times hours\_per\_loci \times overhead$. Assuming 2 hours total integration per pointing, and 20\% overhead, we arrive at 1044 hours for LV and 555 hours for VCC. If we assume that, year round average, a night has 5 useful hours integrating on target, this translates into 209 nights for LV and 111 nights for VCC, where the useful nights for LUCA include gray and dark, spectroscopic conditions and seeing below 1.5 arcsec.

%---------------------------------------------------------------------------------------------------------------------------------------------------------------------------
%---------------------------------------------------------------------------------------------------------------------------------------------------------------------------
\section{\label{sec:instrprop}IFU-6000 Instrument Proposal}

 %% FIGURE:
\begin{figure}
\centering
\includegraphics[width=0.5\textwidth]{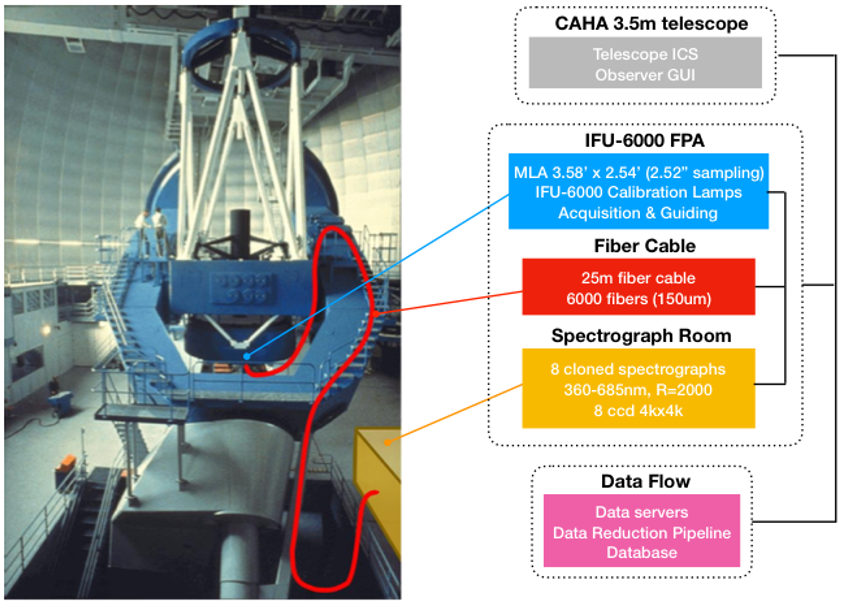}
\caption{
Schematic view of the IFU-6000 instrument at the CAHA 3.5m telescope.
}
\label{fig:systemview6000}
\end{figure}

The main IFU-6000 instrument parameters are listed in table~\ref{table:ifu6000parms}. In the following, we describe each of the instrument systems, including the CAHA 3.5m telescope.

\begin{center}
\begin{table}
\centering
\begin{tabular}{l | l } 
\hline
 IFU-6000 Parameter & As-designed  \\ 
 \hline
Telescope f/ratio & f/10 \\
IFU injection & hex Microlens Array (MLA) \\
FoV diameter per fibre (arcsec) & 2.52 \\
Output f/ratio MLA & f/3.51 \\
Total number of fibres for the IFU & 6000 \\
IFU FoV (arcmin$^2$) ($\sqrt 2$ aspect ratio) & 9.1 ($3.58'\times2.54'$) \\
IFU fill factor & $>95$\% \\
Fibre cable length (meters) & 30 \\
Number of fibres per spectrograph & 750 \\
Number of spectrographs & 8 \\
Wavelength coverage in one arm (nm) & 360-685 \\
Resolution at 500 nm & 2000 \\
Resolution range (over band-pass) & 1440-2740 \\
Projected slit width ($\mu$m) & 150 \\
Slit length (mm) & 162.6 \\
Detector format (number of pixels) & 4k $\times$ 4k \\
Detector pixel size ($\mu$m) & 15 \\
Number of detectors & 8 \\
Fibre core average on CCD (pixels) & 3.73 \\
Spectrograph f/ratio collimator & f/3.43 \\
Spectrograph f/ratio camera & f/1.28 \\
As-designed PSF (maximum r.m.s.) & 12.6 \\
\hline
\end{tabular}
\caption{Summary of main IFU-6000 instrument parameters}
\label{table:ifu6000parms}
\end{table}
\end{center}

\subsection{CAHA 3.5m telescope}

The CAHA 3.5-meter equatorial telescope is located at 2168 m above sea level in Sierra de los Filabres, Almer\'\i a, Spain. It is the largest telescope on continental Europe. The Observatory has excellent facilities, being operated by CAHA, a partnership between the Spanish National Research Council (CSIC) and the Junta de Andaluc\'\i a regional government (JA).

\begin{figure}
\centering
\includegraphics[width=0.5\textwidth]{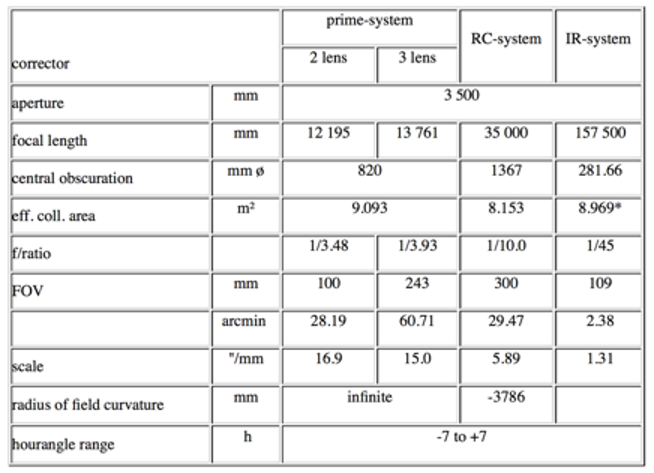}
\caption{
Main optical parameters of the CAHA 3.5m telescope.
}
\label{fig:caha35params}
\end{figure}

Calar Alto is a good astronomical site. Its main properties are similar in many aspects to those of other major observatories. The median seeing is ~0.90$''$ and zenith-corrected values of the moonless night-sky surface brightness are 22.39, 22.86, 22.01, 21.36, and 19.25 mag arcsec$^{-2}$ in U, B, V, R and I, which indicates that Calar Alto is a particularly dark site for optical observations up to the I-band. The fraction of astronomical useful nights at the observatory is ~70\%, with ~30\% of photometric nights. See S\'anchez et al. (2007, PASP, 119, 1186) for all site characterization details.
Table in figure.~\ref{fig:caha35params} displays the main optical parameters of the CAHA 3.5m telescope. Our large IFU will be placed at the focal plane of the Ritchey-Chr\'etien (RC) system focus.

 %% FIGURE:

Our Team has been in close contact with the CAHA staff to follow up their upgrades and feedback regarding the telescope control system (ICS) and had several visits and meetings to evaluate the assessment of the LUCA needs and specifications for the Focal Plane Assembly, specifications and logistics for the fibre cable and spectrograph room. We also had access to the Zemax optical model of the CAHA 3.5m telescope, which was necessary to perform the optical design of the spectrograph.

\subsection{IFU-6000 Focal Plane Assembly}

\begin{figure}
\centering
\includegraphics[width=0.5\textwidth]{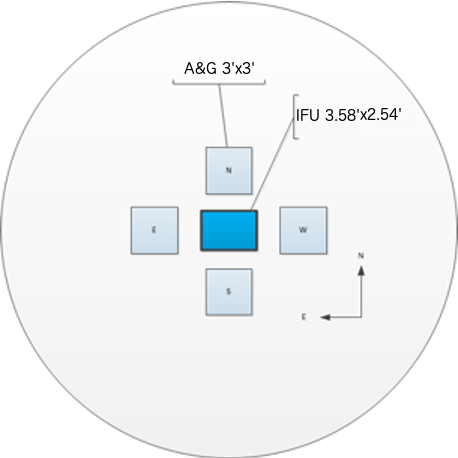}
\caption{
Frontal view of the IFU-6000 Focal Plane Assembly at the 3.5 m RC focus. The large science IFU is surrounded by 4 A\&G cameras. The outer circle has a diameter of 300 mm, i.e. the entire FoV of the RC focus.
}
\label{fig:ifu6000FPA}
\end{figure}

The nominal Focal Plane of the RC focus of the 3.5 m telescope has a practicable circular FoV of $29.47’$ (300 mm) diameter. The large science IFU with a FoV of $3.58’\times2.54’$ will be placed on the optical axis at the center of the focal plane, and four Acquisition \& Guiding (A\&G) cameras, each with a FoV of $3’\times3’$, will be located around at the East, North, West, and South locations. See figure~\ref{fig:ifu6000FPA} schematic view of the Focal Plane Assembly (FPA) as seen from above.

%% FIGURE:

The IFU-6000 FPA will be attached to the 3.5 m A\&G box. Inside the telescope A\&G box we will place the Calibration Unit, by using the same space that currently uses the PMAS Calibration Unit. We will also allow switching from IFU-6000 to the CARMENES high-res spectrograph by moving a mirror into the telescope beam. The PMAS Guiding System can be removed completely, which will allow having more space inside the A\&G box. A detailed technical assessment of the 3.5 m A\&G box in this new context must be performed, so far we do not find any critical or risky aspects.

\begin{figure}
\centering
\includegraphics[width=0.5\textwidth]{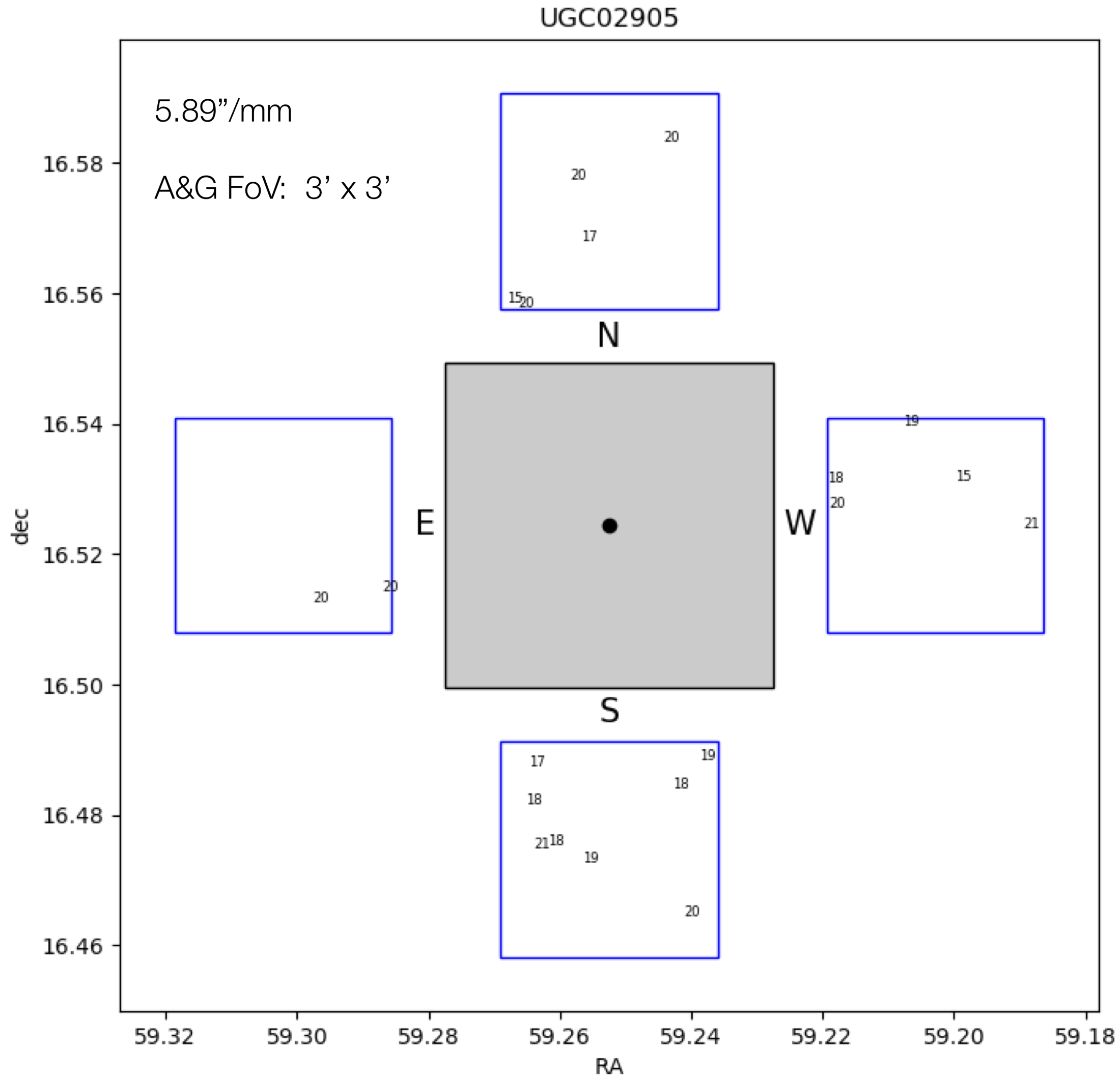}
\caption{
Schematic view of the 20 Gaia guide stars available for the example galaxy UGC02905. The central gray box represents the IFU and the four outer blue boxes correspond to the four auto-guide cameras.
}
\label{fig:guidestar2}
\end{figure}

Acquisition \& Guiding:

The hardware details of the LUCA A\&G camera system can be found in the LUCA Proposal; the paper version of the full Feasibility Study, deposited in the library of the IAA-CSIC, with reference number FT-43(I) and FT-43(II). We use the Gaia database through the \texttt{astroquery} interface to generate sky fields with stars. The FoV to search for available guiding stars at any of the 4 locations on the sky is $3’\times3’$ (see figure~\ref{fig:guidestar2}). 

 %% FIGURE:

\begin{figure}
\centering
\includegraphics[width=0.5\textwidth]{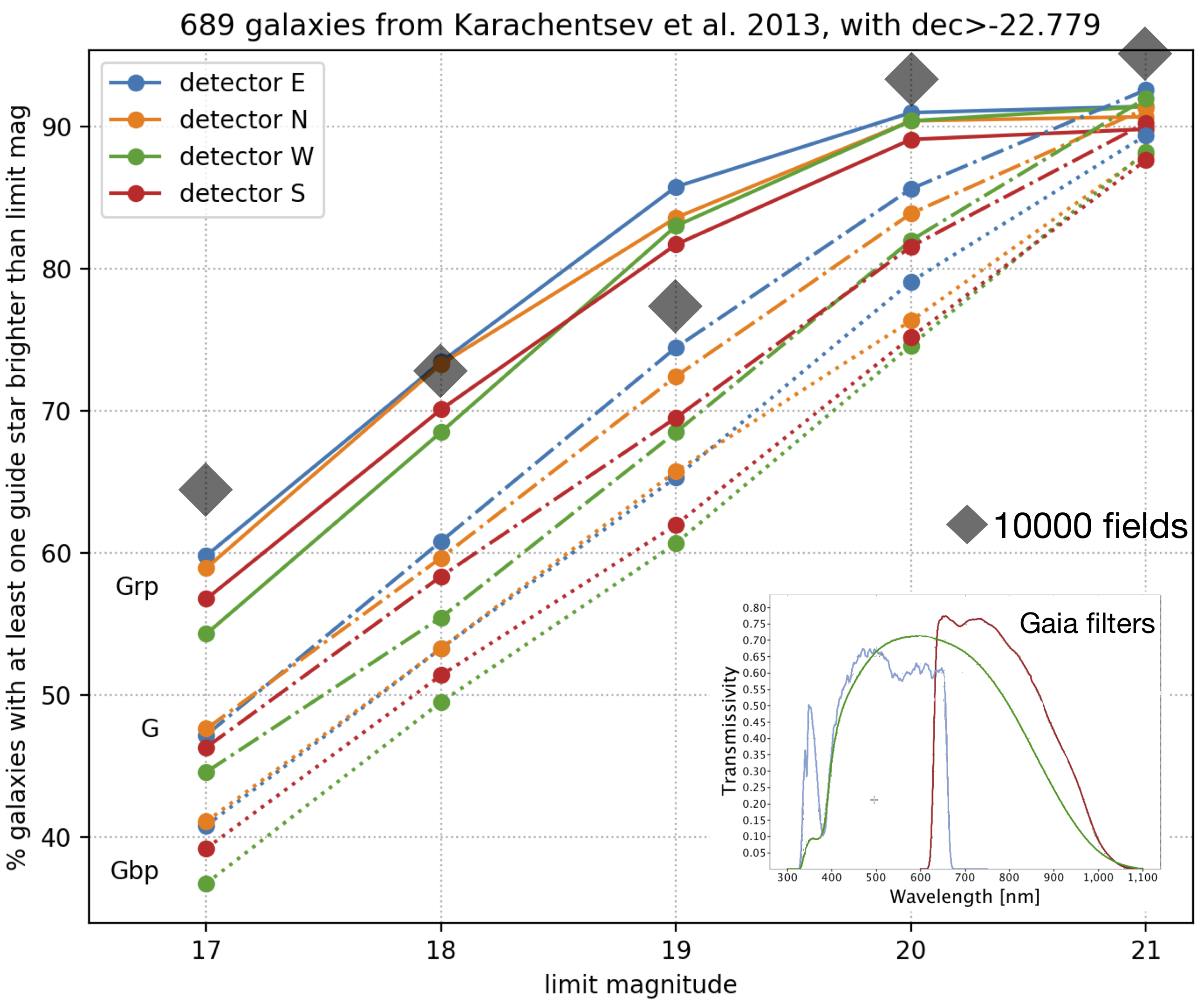}
\caption{
To study the availability of stars for guiding, we generate random locations on a sphere and select 10,000 fields with declination above -22.78 degrees (so that maximum airmass $<2$), see text. 
}
\label{fig:guidestar1}
\end{figure}

To study the availability of stars for guiding for the entire LUCA galaxy sample, we generate random locations on a sphere and select 10,000 fields with declination above -22.78 degrees (so that maximum airmass $<2$). We then count the stars brighter than a given magnitude (Gaia \textit{phot-g-mean-mag}). The gray large squares in figure~\ref{fig:guidestar1} show the percent number of these fields with at least one guide star brighter than a given magnitude. The lines show the actual number of stars for the 689 galaxies in the Karachentsev et al. (2013) LV catalog. The four colors are for the four guide cameras, while the three different line types indicate the values in the three Gaia filters shown in the inset.

%% FIGURE:

\subsection{IFU-6000 Fiber System}

IFU-6000 will furnish the telescope with a powerful short wavelength (360 nm – 680 nm) integral field spectroscopic facility, wherein an optical fiber integral field unit at the Cassegrain focus of the telescope will feed a suite of duplicate spectrographs within a dedicated environmentally controlled enclosure sited on the observatory dome floor, to the east of the equatorial plinth. The IFU will cover 9 square arcminutes of sky with 6000-spaxel, $2.5''$ sampling. A design study was undertaken to examine all relevant aspects of the fiber system, outlining baseline requirements and identifying the most technologically appropriate, low-risk solutions in each case. The study encompassed the following aspects:

\begin{itemize}
\item Fiber selection: transmission, fiber numerical aperture, focal ratio degradation, fiber core size, fiber clad size, fiber buffer material and size, batch variation,
\item Fiber cable design: functional requirements, planetary stranding, 
\item Cable production: furcation tube, installing the fibers into the furcation tube, the tensile element, the stranding production line, conduits, strain relief boxes,
\item Constraints and specific considerations for the cable scheme, 
\item Cable terminations: IFU input (chosen field format and IFU input hardware and optics) and IFU output – slit assemblies,
\item Test strategy,
\item Timeline and Cost estimates.
\end{itemize}

 %% FIGURE:
\begin{figure}
\centering
\includegraphics[width=0.5\textwidth]{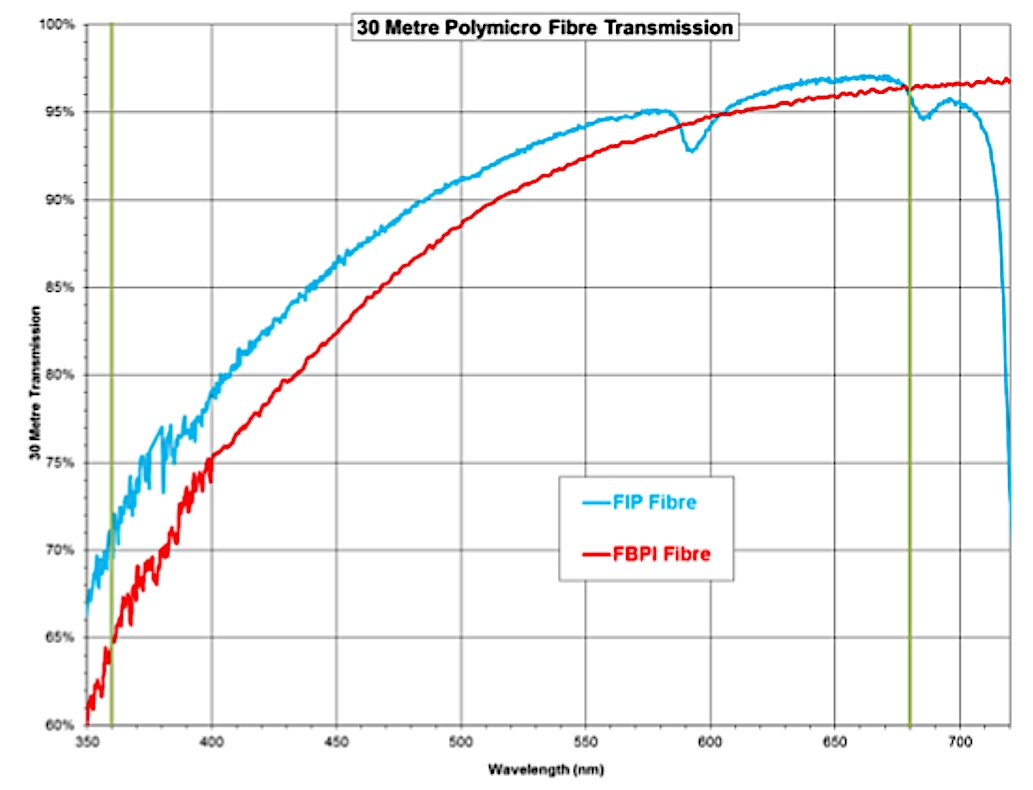}
\caption{
Percent transmission for 30 meter sections of FBPI versus FIP Polymicro fibre from Molex. The desired bandwidth of 360 nm to 680 nm is bounded by the green lines.
}
\label{fig:fibertrans}
\end{figure}

In figure~\ref{fig:fibertrans} we show percentage transmission calculated for 30 meter sections of FBPI versus FIP Polymicro fiber from Molex. The desired bandwidth of 360 nm to 680 nm is bounded by the green lines. With the exception of a small absorption feature centred around 592 nm, the FIP fiber shows a consistently higher transmission than the FBPI product, with the greatest gains evident towards the UV-end of the plots.

{\em IFU with microlenses vs. bare fibres: }

In the simplest case, an IFU can be constructed with bare fibers at both input and output. Here, the F/in at the focal plane is coupled directly into the fiber without modification. Since the cores at the bare fiber input, even if packed in direct contact, will be separated by the cladding and buffer thickness (if retained) there will be a correspondingly low fill factor. Microlenses can be employed to both increase the fill factor and modify the incident F/$\#$. Microlens schemes can be tailored either to feed telescope pupil images or true sub-images from the focal plane into the fibers. Pupil images emerging from the fiber outputs at the pseudo-slit result in field images at the stop of the spectrograph, and vice-versa. While variation in field images at the slit will directly correspond to image variation on the detector, field image variation at the spectrograph stop affects the PSF on the detector. Each solution has its specific advantages and disadvantages and they were studied to determine which is most suited to this application. After undertaking a detailed design trade-off we have chosen to adopt a pupil imaging microlens solution utilising an air spaced microlens doublet array (MLA) at the input, i.e. the AAO-Macquarie “MLA-design”. Although there are additional complications with a doublet scheme (specifically optimal co-registration of the two microlens arrays) the two lenslet scheme still emerges as the optimal solution. Microlenses at the input furnish a fill factor of $91\%$ if configured as contacting circular lenslets, and theoretically almost $100\%$ with hexagonal lenslets. 

%% FIGURE:
\begin{figure}
\centering
\includegraphics[width=0.5\textwidth]{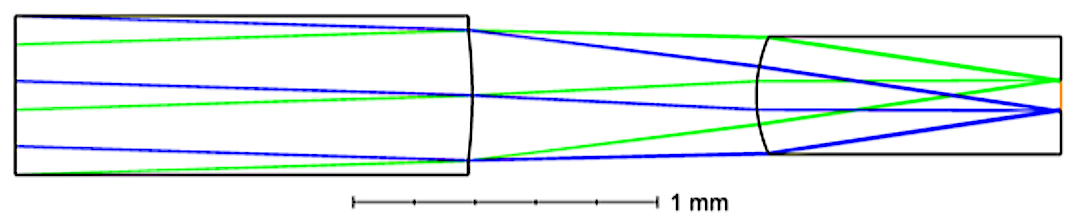}
\caption{
Input microlens system.
}
\label{fig:inputmicrolens}
\end{figure}

\begin{figure}
\centering
\includegraphics[width=0.5\textwidth]{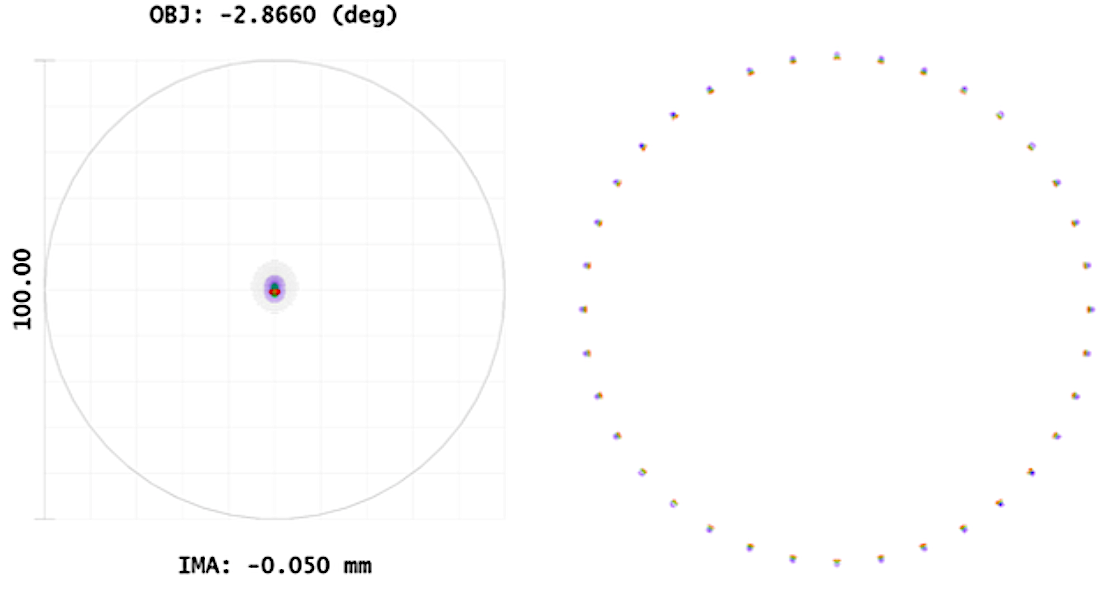}
\caption{
Left: PSF on the fibre core edge of the input microlens system; the circle has the diameter of the fibre core. Right: Aberration FRD of the output beam of the input microlens system.
}
\label{fig:psffibre}
\end{figure}

Figure~\ref{fig:inputmicrolens} shows a design of input microlenses doublet. The focal plane of the telescope is located on the planar face of the first microlens array, with lenslets replicated on the second surface. With the relatively slow F/10 speed of the CAHA 3.5 m secondary that will be used, we are able to avoid fore-optics for field magnification \& telecentric correction. A small residual non-telecentricity of the telescope can be corrected by tuning relative lenslet pitches in the doublet array. The image quality at the edge of the fiber core is shown in the figure~\ref{fig:psffibre} (left panel). This is the only place where the image quality matters. Poor image quality in the center for example would have no consequences since all the light still launches into the fiber core, while the blurring of the PSF at the edge creates losses of light by vignetting. Over-sizing the fibers to avoid losses of light seems an evident solution at first sight but it increases the \'etendue of the system, so the spectrographs must be correspondingly larger and more expensive. 

%% FIGURE:

The right panel of figure~\ref{fig:psffibre} shows the degradation of the beam entering the fiber. That degradation is an average FRD. Both the image qualities on the fiber core and of the beam edge are excellent. The most critical problem of the IFU-6000 is that the seeing is much smaller than the IFU sampling. A typical seeing at Calar Alto is $0.9''$ (median) while the sampling of our IFU is $2.52''$. In principle, the width of the PSF on the IFU should be at least twice the width of the spatial sampling to fulfil the Nyquist sampling criterion, so $6.5''$ for IFU-6000. This would however significantly reduce the resolution and is not that easy to achieve. For example, a defocus of the telescope does not give a Gaussian PSF. It reduces the problem but does not solve it. While the Nyquist sampling criterion is for the ability to precisely reconstruct the intermediate intensities between fibres or microlenses assuming perfect measurements by the spectrograph, it is also valid for the different kinds of measurement errors introduced by the fibre system with or without microlenses. Understanding the consequences on the measurements of all the options is important to determine the best option and calculate its specifications.

{\em Fiber cross-talk along the slit:}

\begin{figure}
\centering
\includegraphics[width=0.5\textwidth]{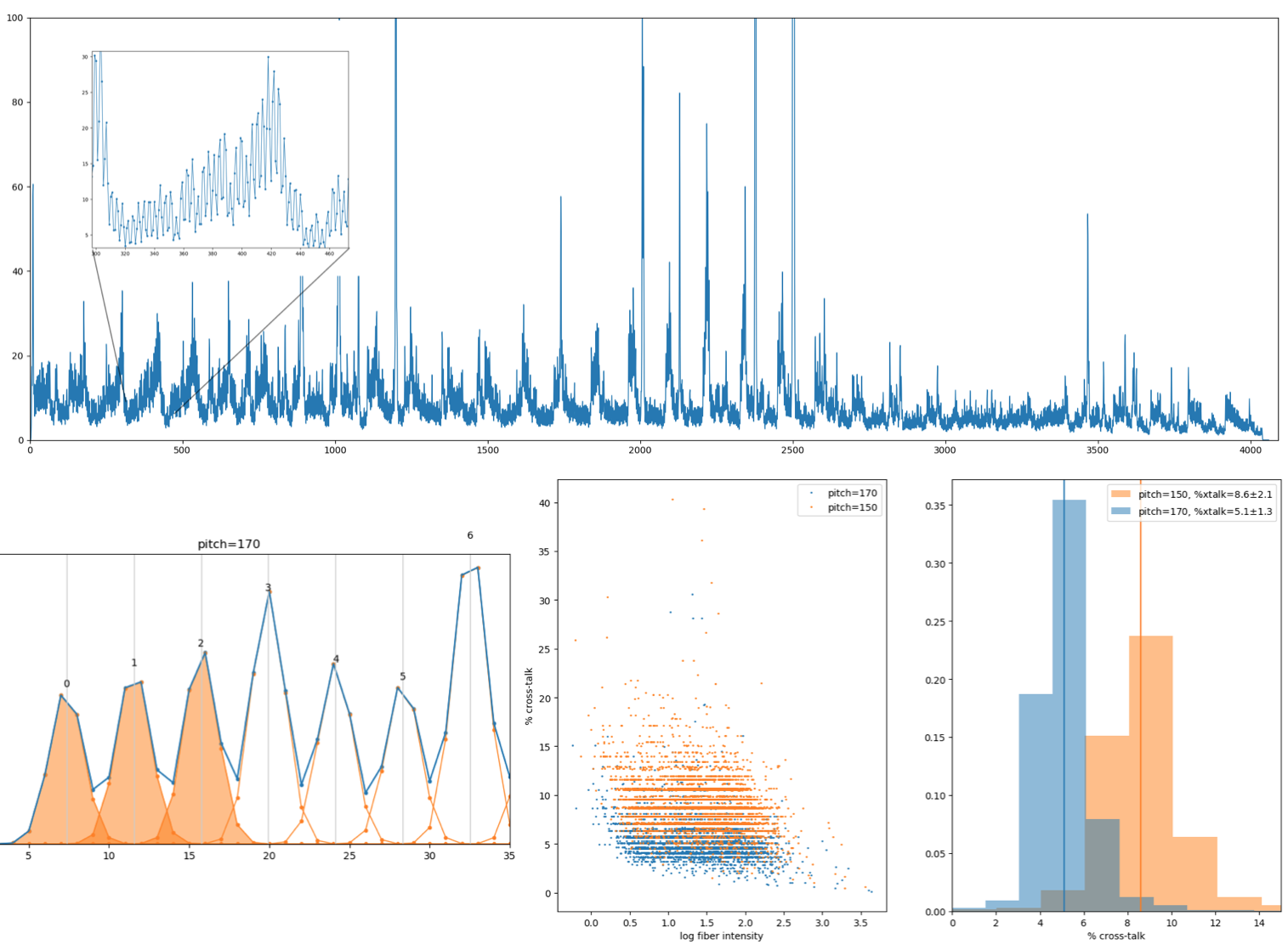}
\caption{
Section along the spatial dimension in the CCD and inset with zoom in where each individual fiber profile can be seen. Bottom left shows the profile fits to extract the flux in each fiber, and the corresponding cross-talk computed and plotted in the bottom right. 
}
\label{fig:crosstalk}
\end{figure}

A single IFU pointing generates a spectral-spatial 2-D image in the detector, with fibers arranged along the spatial direction. A slice of this image along the spatial direction is shown on the top plot in figure~\ref{fig:crosstalk}, magnified in the inset, where the single fiber PSF can be seen. A PSF model (e.g., a Gaussian) is fitted to each fiber PSF as shown in the bottom-left figure (in orange in figure~\ref{fig:crosstalk}), and the cross-talk is defined as the fraction of the PSF that falls below neighboring fibers (darker orange wings). The two bottom-right plots in the figure give the values of cross-talk extracted for two different values of the fiber pitch along the slit (the peak-to-peak separation of fibers along the slit) with values of 150 and 170 microns. We set-up a requirement of a $2\%$ contamination as it was adopted in the AAO spectrograph design.

%% FIGURE:

\subsection{Spectrograph Slit Assembly}

\begin{figure}
\centering
\includegraphics[width=0.5\textwidth]{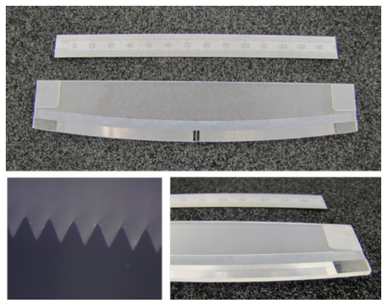}
\caption{
The monolithic v-groove array prototype at CfAI, Durham. Cut from a single piece of borosilicate glass, this v-groove array exhibits extremely high tolerances, high thermal stability, low profile (3 mm height).
}
\label{fig:slit_v_grooves}
\end{figure}

Slit Units comprising linear fiber have been made for numerous instruments. These range from simple parallel schemes to complex, curved / non-telecentric / off-axis geometries. Durham University CfAI has a lot of experience in this area (GMOS \& IMACS IFUs, FMOS, DESI). A slit unit is typically built from individual sub-blocks, with necessary gaps, where the pointing is approximated in parallel fiber sets. However a more advantageous monolithic v-groove array and assembly procedure has recently been prototyped at CfAI, which will be applied to this instrument (see figure~\ref{fig:slit_v_grooves}). Cut from a single piece of borosilicate glass, this v-groove array exhibits true fiber pointing, no gaps, extremely high tolerances, low FRD, high thermal stability, and low profile ($<3$ mm height). This is a unique innovation that should significantly improve the coupling, quality and stability of the PSF. Hence this solution will improve the spectrograph performance compared to more traditional fiber pseudo-slits.

%% FIGURE:

\subsection{IFU-6000 Spectrograph System}

To meet the science requirements of LUCA we proposed building a new IFU fiber-fed spectrograph (IFU-6000) at the CAHA 3.5m telescope that would allow observing in the optical spectral range 360-700 nm with resolution 2000. This IFU will have a unique large FoV of 9 arcmin$^2$ with 6000 fibers and a spatial sampling of $2.5''$ on the sky.

Our collaborators at AAO-Macquarie and Winlight System have performed a design study for the IFU-6000 spectrograph (their detailed studies can be found in the full document proposal). Note that only the AAO-Macquarie spectrograph design has included the necessary pre-optics system. Our LUCA team has been in close contact and had the necessary interaction and feedback with both groups during the development of the feasibility study. We also had face-to-face meetings at Winlight Systems in Marseille at the start of the work, and had the visit of Jon Lawrence, Head of Technology of AAO-Macquarie, at the IAA in Granada.

\begin{figure*}
\centering
\includegraphics[width=1.0\textwidth]{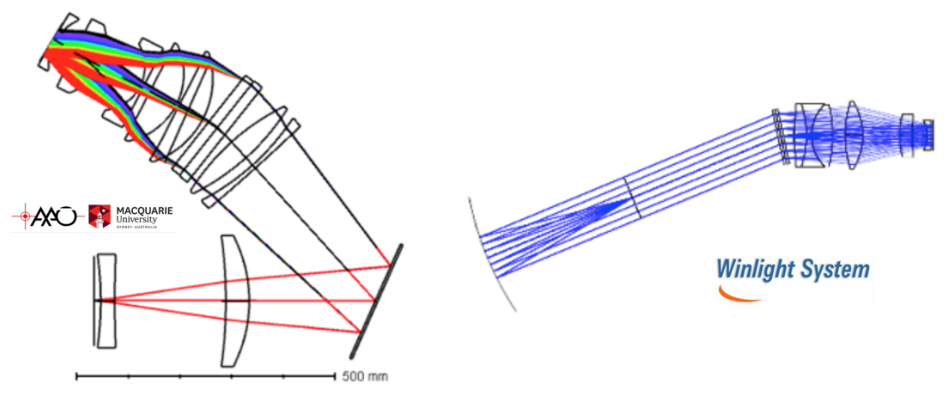}
\caption{
Designs of the IFU-6000 spectrograph. Left: Optical layout for the AAO-Macquarie “MLA design”. This design shows a mirror in the collimator but this would be removed to follow a straight-through layout or be replaced by a dichroic in case a red-arm camera wants to be developed as future upgrade. The slit is flat. Right: Winlight System optical layout of their “DESI 1-arm” design proposal. In this case the collimator is a spherical mirror and the slit is curved. Both designs use a VPH as disperser element.
}
\label{fig:aao_winlight_spec}
\end{figure*}

AAO-Macquarie have developed designs for a baseline version that uses bare fiber injection and a version that uses microlens arrays (MLA). They also have done a compact design (smaller optics that accommodates $25\%$ smaller number of fibers per spectrograph) with lower spectrograph unit cost, but that increases considerably the number of spectrographs, i.e. 13 units vs. 8 units for the baseline version. Here, we have considered the ‘MLA design’ as our preferred option, see figure~\ref{fig:aao_winlight_spec}, as compared to that with bare fibers since it  guarantees a nearly perfect homogeneity in $S/N$ over the entire FoV (hard to get with bare fiber IFU, which only have a $54\%$ filling factor as compared to $>95$\% obtained with an hexagonal MLA). In addition the ‘MLA design’ doesn’t need any additional pre-optics. This saves cost and complexity. The design with a bare-fibre IFU needs a pre-optics, which is a doublet and 2 singlet lenses as an optical relay converting the f/10 beam from the CAHA 3.5m telescope focus to give f/3.5 at the intermediate focus where the fibre IFU is located. Three of the surfaces are aspheric.
 
%% FIGURE:

The Winlight proposed design for the IFU-6000 spectrograph is an update of the DESI blue optical configuration (fiber n.a.=0.14, curved slit and spherical collimator mirror), the main difference being the addition of a lens (see figure \ref{fig:aao_winlight_spec}). The asphere is also moved from the field lens to this new lens. The collimator is also slightly updated.

The opto-mechanical design for both AAO-Macquarie and Winlight spectrographs are described in detail in the full document proposal. Note that the AAO-Macquarie design does not require any moving parts. In the case of Winlight System their design requires two moving elements: the collimator Hartman doors and the shutter, which is placed at the collimated beam. In the case of the AAO-Macquarie the shutter is behind the spectrograph camera and attached to the CCD camera.

\begin{figure*}
\centering
\includegraphics[width=0.5\textwidth]{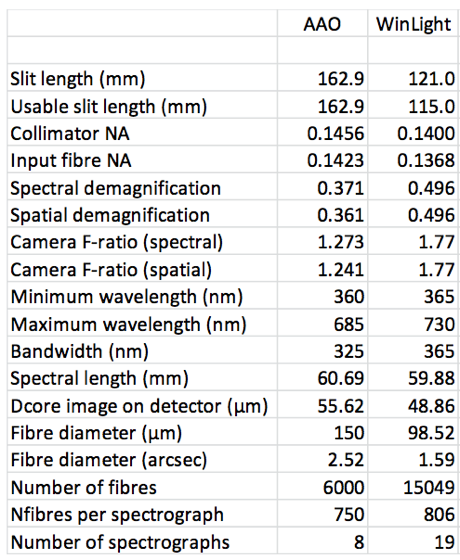}
\caption{
AAO-Macquarie vs. Winlight spectrograph parameters.
}
\label{fig:AAOvsWinlight}
\end{figure*}

Below we collect a number of reasons that justify the preference of the AAO-Macquarie proposal over that from Winlight System, after taking in consideration science and technical requirements:

\begin{itemize}
\item To get a spectral resolution of R=2000 at 500 nm and a FoV of 9 arcmin$^2$ on the sky, the Winlight fibers would need to be $1.59''$ wide instead of $2.52''$. Then 15,050 fibers would be needed to map the entire FoV, and hence 19 Winlight spectrographs compared to 8 AAO-Macquarie spectrographs with a total of 6000 fibers (see Figure \ref{fig:AAOvsWinlight}),
\item The slit of the Winlight spectrograph has its pupil behind it. This creates collision between the fibers at the back of the slit so more space may be needed. The AAO-Macquarie design has the pupil in front of the slit so there is more space behind the slit and no losses,
\item The Winlight slit is on-axis right in the path of the light reflected by the mirror. This has similar effect than a spider in a Schmidt camera. Apart from vignetting some light, probably 1\% to a few percent, it generates diffraction spikes, which may also affect the PSF,
\item The light density in arcsec$^2$/pixel is smaller on the Winlight spectrographs than on the AAO-Macquarie's by a factor of 1.84. This means  more detector noise (read-out + dark current) on the Winlight spectra. The $S/N$ ratio degradation will be at its maximum at the blue-end,
\item AAO-Macquarie slit is flat as compared to the curved Winlight slit. The latter won’t allow having microlenses on it in case we do need them to optimize the spectrograph performance,
\item There is not much difference between the two designs in terms of average transmission of the optics,
\item The spectral bandwidth of AAO-Macquarie (360-685 nm) vs. Winlight (365-730 nm) offers an advantage thanks to its bluer minimum wavelength. The AAO-Macquarie 685 nm red limit also guarantees to observe all main relevant spectral features. There is no  significant gain in terms of science performance going up to 730 nm,
\item Despite the fact that the cost of a Winlight spectrograph is about 40\% cheaper that one AAO-Macquarie unit, a total of 19 Winlight units are needed as compared to the 8 AAO-Macquarie’s, which will largely increase the total cost of the project by a factor of 2 only due to the spectrographs, ignoring the increase of cost by the needed of additional 11 CCDs.
\end{itemize}

%% FIGURE:

See below a chart with the main features of the IFU-6000 spectrograph adopting the AAO-Macquarie “MLA desing” as our preferred option. 

%% FIGURE:
\begin{figure*}
\centering
\includegraphics[width=1.0\textwidth]{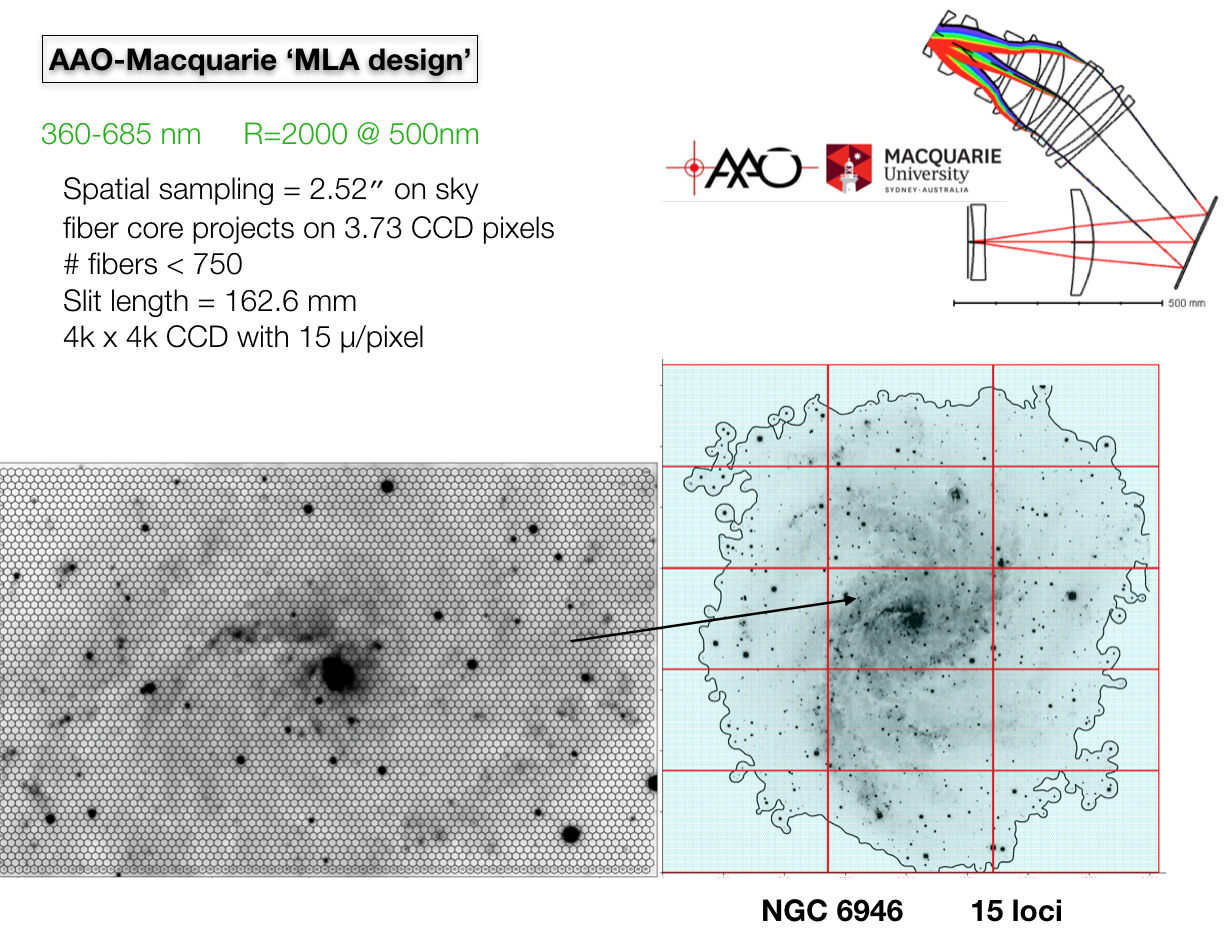}
\caption{
Global summary of the AAO-Macquarie spectrograph design, parameters, and view of an example galaxy.
}
\label{fig:AAOdesign}
\end{figure*}

The AAO-Macquarie design should have been optimized and developed further in the next phase of the project to complete its Final Design Review, if LUCA were approved.

We were considering the scientific e2v CCD Sensor 4096$\times$4096 pixels (CCD231-84 back illuminated) for the spectrographs. This CCD meets our science requirements. The cryostat would have been that used in the AAO TAIPAN and Hector spectrographs.

%---------------------------------------------------------------------------------------------------------------------------------------------------------------------------
\section{\label{sec:context}LUCA in the worldwide context}

\begin{figure}
\centering
\includegraphics[width=0.5\textwidth]{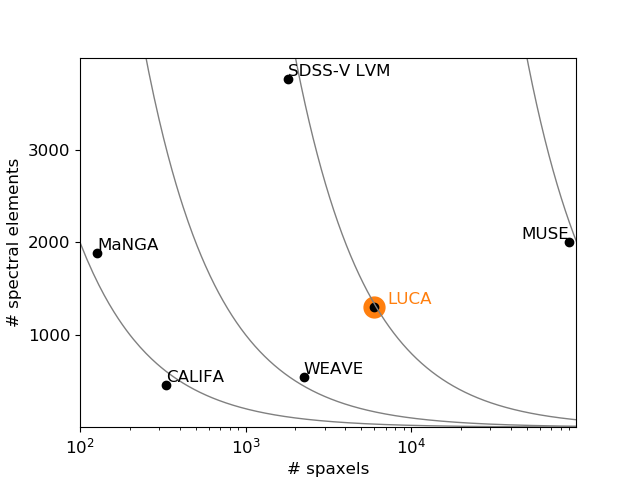}
\caption{
Number of spectral elements versus number of spatial elements for the worldwide current major IFU instruments / surveys. The full lines indicate different values of the constant ratio of the two parameters.
}
\label{fig:etendue-1}
\end{figure}

We compare LUCA with the most competitive existing and upcoming IFUs: MaNGA at the SDSS 2.5m telescope, CALIFA at the CAHA 3.5m telescope, SDSS-V Local Volume Mapper at the SDSS 2.5m telescope, WEAVE at the WHT 4.2m telescope, and MUSE at the VLT 8.2m telescope. Figure~\ref{fig:etendue-1} displays all these IFUs given their total number of pixels and total number of spectral resolution elements. The solid lines indicate IFUs with the same ratio of $\# spaxels$ and $\# spectral$ elements.

 %% FIGURE:

\begin{figure}
\centering
\includegraphics[width=0.5\textwidth]{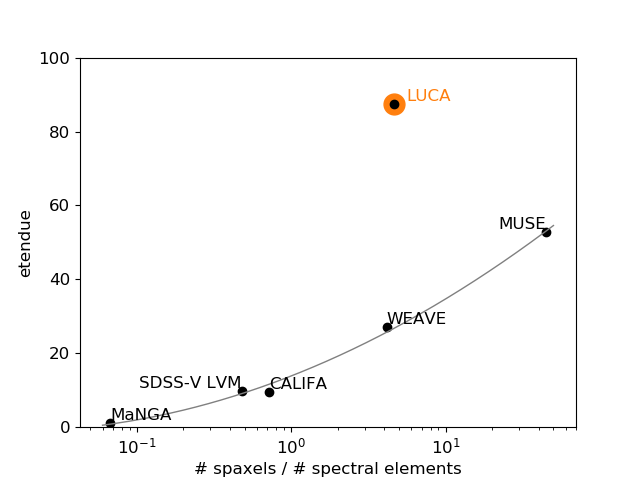}
\caption{
Etendue versus the ratio of spatial to spectral elements for the current major IFU instruments/surveys. LUCA stands out high above all other projects in \'etendue. Overall, the systems seem to lie along a curve given by a "golden-law". This is an interesting result that needs to be explored further. The full line is a second order poly-fit.  
}
\label{fig:etendue-2}
\end{figure}

The competitiveness of an IFU to perform the proposed survey of the Local Volume galaxies relies on its capability of being efficient in mapping a given area of the sky that covers each galaxy down to some limiting surface magnitude. In this regard the IFU \'etendue is the key parameter that drives the speed of the survey. Figure~\ref{fig:etendue-2} compares the \'etendue and the ratio of $\# spaxels$ to $\# spectral$ elements of LUCA with the other IFUs. It can be clearly seen that LUCA has by far the largest \'etendue, being almost one order of magnitude larger than the SDSS-V LVM. Interestingly, we find that the existing and planned IFUs obey some “golden-law” as shown by the solid curve. LUCA appears many sigmas away from this “law”, which demonstrates its outstanding unique and competitive niche for Integral Field Spectroscopy surveys.

 %% FIGURE:

The SDSS-V Local Volume Mapper is planned to start observations beyond 2020. The details can be found in arXiv:1711.03234, and it intends to perform IFU mapping of the Milky Way, LMC, M31, M33, and other galaxies out to 5 Mpc. They plan to do sparse spatial sampling of these galaxies, and also statistical samples of HII regions at 20 pc resolution in M31 and $\sim50$ pc in other galaxies. 

SDSS-V LVM is complementary to LUCA since it does not intend to perform an entire IFU spatial mapping of all galaxies within our Local Volume. On the other hand, as seen above, the \'etendue of LUCA is almost one order of magnitude larger than that of the SDSS-V LVM! Thus, {\bf LUCA is a unique IFU spectroscopic facility that can offer the worldwide astronomical community an invaluable legacy, which will be fundamental in our understanding of galaxy formation and evolution, and will result key for follow-up detailed studies in the ELT era.}

%---------------------------------------------------------------------------------------------------------------------------------------------------------------------------
\section{\label{sec:conclus}Conclusions and remarks}

This document summarises the Feasibility Study of a new fiber-fed IFU spectrograph (IFU-6000) at the CAHA 3.5m telescope. IFU-6000 has a huge \'etendue and large number of spatial / spectral elements compared to other existing and upcoming worldwide IFS. The new instrument facility would have allow astronomers to complete the LUCA Local Volume and Virgo surveys, which aim to map our universe neighborhood in 3D with an unprecedented spatial resolution (better than 100 pc) to unveil the sub-grid physics of galaxy formation.
The work included in this report provides an overview of the LUCA science program, science specifications and requirements of the instrument, science survey strategy, a technical description of the instrument technical proposal, and discuss the project competitiveness worldwide.
We have shown technical solutions and feasibility for all instrument systems. There are no high-risk items identified because this project will follow closely other very similar projects built by the LUCA team using the same vendors for which there is a good track record. This is especially important for the serial production of the cloned spectrographs. 

The project, being ready to start its Preliminary Design phase, was canceled unilaterally by the IAA-CSIC. We do hope that the LUCA Feasibility Study effort could benefit the astronomical community and those interested groups that aim to implement a similar science and instrumentation program in 4-m class telescopes.

\section{\label{sec:acknow}Acknowledgements}

The LUCA Feasibility Study has been performed and led by the LUCA team at the IAA-CSIC in collaboration with AAO-Macquarie University, Durham University, and Winlight System. The LUCA team at IAA-CSIC is integrated by Francisco Prada (Principal Investigator, Project Manager, Instrument Scientist), Enrique P\'erez (Project Scientist), Justo S\'anchez (Instrument Engineer), and Jos\'e Miguel Ib\'a\~nez (Software and Control Engineer). Graham Murray at Durham University contributed with the fiber system and slit assembly work; Robert Content and Jon Lawrence at AAO-Macquarie with the pre-optics and spectrograph system, and the company Winlight System provided the spectrograph design for their solution.

A total of 243,300 euro was required for its accomplishment. Out of the 100,000 euro provided by CAHA through the funding given by the Junta de Andaluc\'\i a and managed by the University of Almer\'\i a, the project spent 76,300 euro to pay the consultancy work done by AAO-Macquarie, Durham University, and Winlight System. The human resources devoted by the IAA-CSIC to the feasibility study sum a total of 161,000 euro, which correspond to a total of 2.35 FTE. In addition 6,000 euro was spent by the IAA-CSIC in travel and meetings.

We thank the CAHA staff Gilles Bergond, Jens Helming, Luis Hernández, and Santiago Reinhart for their feedback on the CAHA 3.5m telescope specifications, interfaces and observatory infrastructures. We acknowledge María Balaguer, Head of UDIT at IAA-CSIC, for her comments and recommendations on Project Management, System Engineering and AIV. We also thank Rub\'en Garc\'\i a Benito, Luca Izzo, and Enrique P\'erez-Montero at the IAA-CSIC for useful science discussions, and Johan Comparat at MPE in Garching for help with the exposure time estimations. We want to thank Rosa Gonz\'alez Delgado for her involvement and work in the LUCA Proposal that was submitted to CAHA in the call for new instrumentation. We are grateful to Eric Prieto at LAM in Marseille for his feedback on the Winlight spectrograph design, and Simon Tulloch (QUCAM) for his advice and suggestions on the Detector System and readout electronics.

This research has made use of the SIMBAD database (Wenger et al 2000, A\&AS 143, 9), operated at CDS, Strasbourg, France.

This work has made use of data from the European Space Agency (ESA) mission {\it Gaia} (\url{https://www.cosmos.esa.int/gaia}), processed by the {\it Gaia} Data Processing and Analysis Consortium (DPAC, \url{https://www.cosmos.esa.int/web/gaia/dpac/consortium}). Funding for the DPAC has been provided by national institutions, in particular the institutions participating in the {\it Gaia} Multilateral Agreement.

\end{document}